\newcommand\apjcls{1}
\newcommand\aastexcls{2}
\newcommand\othercls{3}

\newcommand\papercls{\aastexcls}
\documentclass[tighten, times, twocolumn]{aprilis_aastex62}  

\if\papercls \apjcls
\usepackage{apjfonts}
\else\if\papercls \othercls
\usepackage{epsfig}
\usepackage{margin}
\usepackage{times}
\fi\fi
\usepackage{ifthen}
\usepackage{natbib}
\usepackage{amssymb, amsmath}
\usepackage{appendix}
\usepackage{etoolbox}
\usepackage[T1]{fontenc}
\usepackage{paralist}

\if\papercls \apjcls
\newcommand\aas{\ref@jnl{AAS Meeting Abstracts}}
\newcommand\dps{\ref@jnl{AAS/DPS Meeting Abstracts}}
\newcommand\maps{\ref@jnl{MAPS}}
\else\if\papercls \othercls
\usepackage{astjnlabbrev-jh}
\fi\fi

\if@Spr@basic@refstyle%
\if@Numbered@refstyle%
  \usepackage[numbers,sort&compress]{natbib}%
\else%
 \usepackage[authoryear]{natbib}%
\fi%
  \def\bibfont{\reset@font\fontfamily{\rmdefault}\normalsize\selectfont}%
\fi%

\if@Mathphys@numrefstyle%
  \usepackage[numbers,sort&compress]{natbib}%
  \def\bibfont{\reset@font\fontfamily{\rmdefault}\normalsize\selectfont}%
\else%
\if@Mathphys@ayrefstyle%
 \usepackage[authoryear]{natbib}%
  \def\bibfont{\reset@font\fontfamily{\rmdefault}\normalsize\selectfont}%
\fi\fi%
\if@APS@refstyle%
  \usepackage[numbers,sort&compress]{natbib}%
  \def\bibfont{\reset@font\fontfamily{\rmdefault}\normalsize\selectfont}%
\fi%
\if@Vancouver@refstyle%
\if@Numbered@refstyle%
  \usepackage[numbers,sort&compress]{natbib}%
\else%
 \usepackage[authoryear]{natbib}%
\fi%
  \def\bibfont{\reset@font\fontfamily{\rmdefault}\normalsize\selectfont}%
\fi%
\if@APA@refstyle%
\if@Numbered@refstyle%
  \usepackage[natbibapa]{apacite}%
\else%
  \usepackage[natbibapa]{apacite}%
\fi%
  \def\refdoi#1{\urlstyle{rm}\url{#1}}%
  \AtBeginDocument{%
  }%
  \def\bibfont{\reset@font\fontfamily{\rmdefault}\normalsize\selectfont}%
\fi%
\if@Chicago@refstyle%
\if@Numbered@refstyle%
  \usepackage[numbers,sort&compress]{natbib}%
\else%
 \usepackage[authoryear]{natbib}%
\fi%
  \hypersetup{urlcolor=black,colorlinks=false,pdfborder={0 0 0}}
  \def\bibfont{\reset@font\fontfamily{\rmdefault}\normalsize\selectfont}%
\fi%
\if@Standard@Nature@refstyle%
  \usepackage[numbers,sort&compress]{natbib}%
  \def\bibfont{\reset@font\fontfamily{\rmdefault}\normalsize\selectfont}%
\fi%

\AtBeginDocument{\allowdisplaybreaks}%

\def\eqnheadfont{\reset@font\fontfamily{\rmdefault}\fontsize{16}{18}\bfseries\selectfont}%

\if\papercls \aastexcls
\hypersetup{citecolor=blue, 
            linkcolor=blue, 
            menucolor=blue, 
            urlcolor=blue}  
\else
\usepackage[
bookmarks=true,           
bookmarksnumbered=true,   
colorlinks=true,          
citecolor=blue,           
linkcolor=blue,           
menucolor=blue,           
urlcolor=blue,            
linkbordercolor={0 0 1},  
pdfborder={0 0 1},
frenchlinks=true]{hyperref}
\fi

\if\papercls \othercls

\else

\fi

\providecommand{\adsurl}[1]{\href{#1}{ADS}}

\makeatletter
\patchcmd{\NAT@citex}
  {\@citea\NAT@hyper@{%
     \NAT@nmfmt{\NAT@nm}%
     \hyper@natlinkbreak{\NAT@aysep\NAT@spacechar}{\@citeb\@extra@b@citeb}%
     \NAT@date}}
  {\@citea\NAT@nmfmt{\NAT@nm}%
   \NAT@aysep\NAT@spacechar\NAT@hyper@{\NAT@date}}{}{}

\patchcmd{\NAT@citex}
  {\@citea\NAT@hyper@{%
     \NAT@nmfmt{\NAT@nm}%
     \hyper@natlinkbreak{\NAT@spacechar\NAT@@open\if*#1*\else#1\NAT@spacechar\fi}%
       {\@citeb\@extra@b@citeb}%
     \NAT@date}}
  {\@citea\NAT@nmfmt{\NAT@nm}%
   \NAT@spacechar\NAT@@open\if*#1*\else#1\NAT@spacechar\fi\NAT@hyper@{\NAT@date}}
  {}{}
\makeatother

\makeatletter
\DeclareRobustCommand{\lowcase}[1]{\@lowcase#1\@nil}
\def\@lowcase#1\@nil{\if\relax#1\relax\else\MakeLowercase{#1}\fi}
\makeatother

\DeclareSymbolFont{UPM}{U}{eur}{m}{n}
\DeclareMathSymbol{\umu}{0}{UPM}{"16}
\let\oldumu=\umu
\renewcommand\umu{\ifmmode\oldumu\else\math{\oldumu}\fi}

\if\papercls \othercls

\else

\fi

\let\oldsim=\sim
\renewcommand\sim{\ifmmode\oldsim\else\math{\oldsim}\fi}
\let\oldpm=\pm
\renewcommand\pm{\ifmmode\oldpm\else\math{\oldpm}\fi}
\newcommand\by{\ifmmode\times\else\math{\times}\fi}

\newbox{\wdbox}
\renewcommand\c{\setbox\wdbox=\hbox{,}\hspace{\wd\wdbox}}
\renewcommand\i{\setbox\wdbox=\hbox{i}\hspace{\wd\wdbox}}

\newcount\timect
\newcount\hourct
\newcount\minct
\newcommand\now{\timect=\time \divide\timect by 60
         \hourct=\timect \multiply\hourct by 60
         \minct=\time \advance\minct by -\hourct
         \number\timect:\ifnum \minct < 10 0\fi\number\minct}

\catcode`@=11

\newcommand\comment[1]{}

\newcommand\commenton{\catcode`\%=14}

\renewcommand\math[1]{$#1$}
\newcommand\mathshifton{\catcode`\$=3}

\let\atab=&
\newcommand\atabon{\catcode`\&=4}

\let\oldmsp=\sp
\let\oldmsb=\sb
\def\sp#1{\ifmmode
           \oldmsp{#1}%
         \else\strut\raise.85ex\hbox{\scriptsize #1}\fi}
\def\sb#1{\ifmmode
           \oldmsb{#1}%
         \else\strut\raise-.54ex\hbox{\scriptsize #1}\fi}
\newbox\@sp
\newbox\@sb
\def\sbp#1#2{\ifmmode%
           \oldmsb{#1}\oldmsp{#2}%
         \else
           \setbox\@sb=\hbox{\sb{#1}}%
           \setbox\@sp=\hbox{\sp{#2}}%
           \rlap{\copy\@sb}\copy\@sp
           \ifdim \wd\@sb >\wd\@sp
             \hskip -\wd\@sp \hskip \wd\@sb
           \fi
        \fi}
\def\msp#1{\ifmmode
           \oldmsp{#1}
         \else \math{\oldmsp{#1}}\fi}
\def\msb#1{\ifmmode
           \oldmsb{#1}
         \else \math{\oldmsb{#1}}\fi}

\def\supon{\catcode`\^=7}

\def\subon{\catcode`\_=8}

\def\supsubon{\supon \subon}

\newcommand\actcharon{\catcode`\~=13}

\newcommand\paramon{\catcode`\#=6}

\comment{And now to turn us totally on and off...}

\newcommand\reservedcharson{ \commenton  \mathshifton  \atabon  \supsubon 
                             \actcharon  \paramon}

\catcode`@=12
\reservedcharson

\if\papercls \apjcls

\else

\fi

\newcommand\SST{{\em Spitzer Space Telescope}}
\newcommand\Spitzer{{\em Spitzer}}
\newcommand\Webb{{\em James Webb Space Telescope}}
\newcommand\JWST{{\em JWST}}
\newcommand\HST{{\em HST}}
\newcommand\chisq{\ifmmode{\chi\sp{2}}\else\math{\chi\sp{2}}\fi}
\newcommand\redchisq{\ifmmode{ \chi\sp{2}\sb{\rm red}}
                    \else\math{\chi\sp{2}\sb{\rm red}}\fi}
\newcommand\Teq{\ifmmode{T\sb{\rm eq}}\else$T$\sb{eq}\fi}
\newcommand\mjup{\ifmmode{M\sb{\rm Jup}}\else$M$\sb{Jup}\fi}
\newcommand\rjup{\ifmmode{R\sb{\rm Jup}}\else$R$\sb{Jup}\fi}
\newcommand\msun{\ifmmode{M\sb{\odot}}\else$M\sb{\odot}$\fi}
\newcommand\rsun{\ifmmode{R\sb{\odot}}\else$R\sb{\odot}$\fi}
\newcommand\mearth{\ifmmode{M\sb{\oplus}}\else$M\sb{\oplus}$\fi}
\newcommand\rearth{\ifmmode{R\sb{\oplus}}\else$R\sb{\oplus}$\fi}

\usepackage{newtxtext}  
\usepackage{listings}
\usepackage{xcolor}

\definecolor{codegreen}{rgb}{0,0.6,0}
\definecolor{codegray}{rgb}{0.5,0.5,0.5}
\definecolor{codepurple}{rgb}{0.58,0,0.82}
\definecolor{backcolour}{rgb}{0.95,0.95,0.92}
\definecolor{ghost}{rgb}{0.95, 0.95, 0.95}

\lstdefinestyle{mystyle}{
    backgroundcolor=\color{backcolour},
    commentstyle=\color{codegreen},
    keywordstyle=\color{magenta},
    numberstyle=\tiny\color{codegray},
    stringstyle=\color{codepurple},
    basicstyle=\ttfamily\footnotesize,
    breakatwhitespace=false,
    breaklines=true,
    captionpos=b,
    keepspaces=true,
    numbers=left,
    numbersep=5pt,
    showspaces=false,
    showstringspaces=false,
    showtabs=false,
    tabsize=2
}

\newcommand\puppies{\textsc{puppies}}
\shorttitle{The Public Photometry Pipeline for Exoplanets}
\shortauthors{P. E. Cubillos}

\begin{document}

\title{The Public Photometry Pipelines for Exoplanets}
\author{Patricio~E.~Cubillos}
\affiliation{Space Resort Institute\footnote{See Appendix \ref{sec:resort}.}, Graz, Austria}
\email{pcubillos@fulbrightmail.org}

\begin{abstract}
Over the past decade, exoplanet atmospheric characterization has
became what some might call the cosmology of astronomy.  In an attempt
to extract and understand the weak planetary signals (a few percent
down to a few tens of ppm times that of their host-star signals),
researchers have developed dozens of idealized planetary atmospheric
models.  Physical interpretations hinge on pretending that we
understand stellar signals (as well behaved mostly temporarily static
spherical cows), as well as planetary signals (as unidimensional
objects, or sometimes quasi-multidimensional objects).  The discovery
of small and cool planets has lead to analyze planetary signals well
below the designed photometric precision of current instrumentation.
The challenge is up there, and keep us busy, so all is well.  Here we
present yet another open-source tool to analyze exoplanet data of
time-series observations.  The {\puppies} code is available via PyPI
(\texttt{pip install exo-puppies}) and conda, the documentation is
located at \href{https://puppies.rtfd.io} {https://puppies.rtfd.io}.

\end{abstract}


\section{Before the Dawn}
\label{introduction}


Without a doubt, the world events that occurred during the past few
years and that culminated during 2021 have been some of the most hard
moments that we all have endured.  Disruptions at the global scale
like these can put an additional mental strain on the population at
all levels.  Now that sufficient time has passed, we can take a more
objective view of these events.  Of course, we are talking
about \#FreeBritney.

Let's recap, after pop-star Britney Spears' was
distraught from a series of difficult years (2006--2008),
her father Jamie Spears put her under a conservatorship, allowing him
to take control of her finances, healthcare, daily routine, and pretty
much every aspect of her life.  Obviously, since a conservatorship
deprives the person of all their liberties, it is imposed only as a
last resort, when a person's decision-making capacity is impaired.
This is why Britney's conservatorship raised so many red flags, she was
a young person that continued working even during the conservatorship,
which lasted thirteen years. In the word of Britney asking the Los
Angeles Superior Court to end the conservatorship (June 23,
2021)\footnote{\href{https://www.npr.org/2021/06/24/1009858617}{https://www.npr.org/2021/06/24/1009858617}}:

``The control he [her dad] had over someone as powerful as me...  It's
embarrassing and demoralizing what I've been through, and that's the
main reason I've never said it openly.  I didn't think anyone would
believe me.
I've been in denial, I've been in shock.  I am traumatized.
I'm so angry, it's insane.
My dad and anyone involved in this conservatorship should be in jail.
I'm telling you this again [addressing Judge B. Penny], so maybe
you can understand the depth and the degree and the damage that they
did to me back then.  I feel bullied, I feel left out and alone.  And
I'm tired of feeling alone.  I deserve to have the same rights as
anybody does.  I just want my life
back.''

\begin{figure}[t]
\centering
\includegraphics[width=\linewidth, clip]{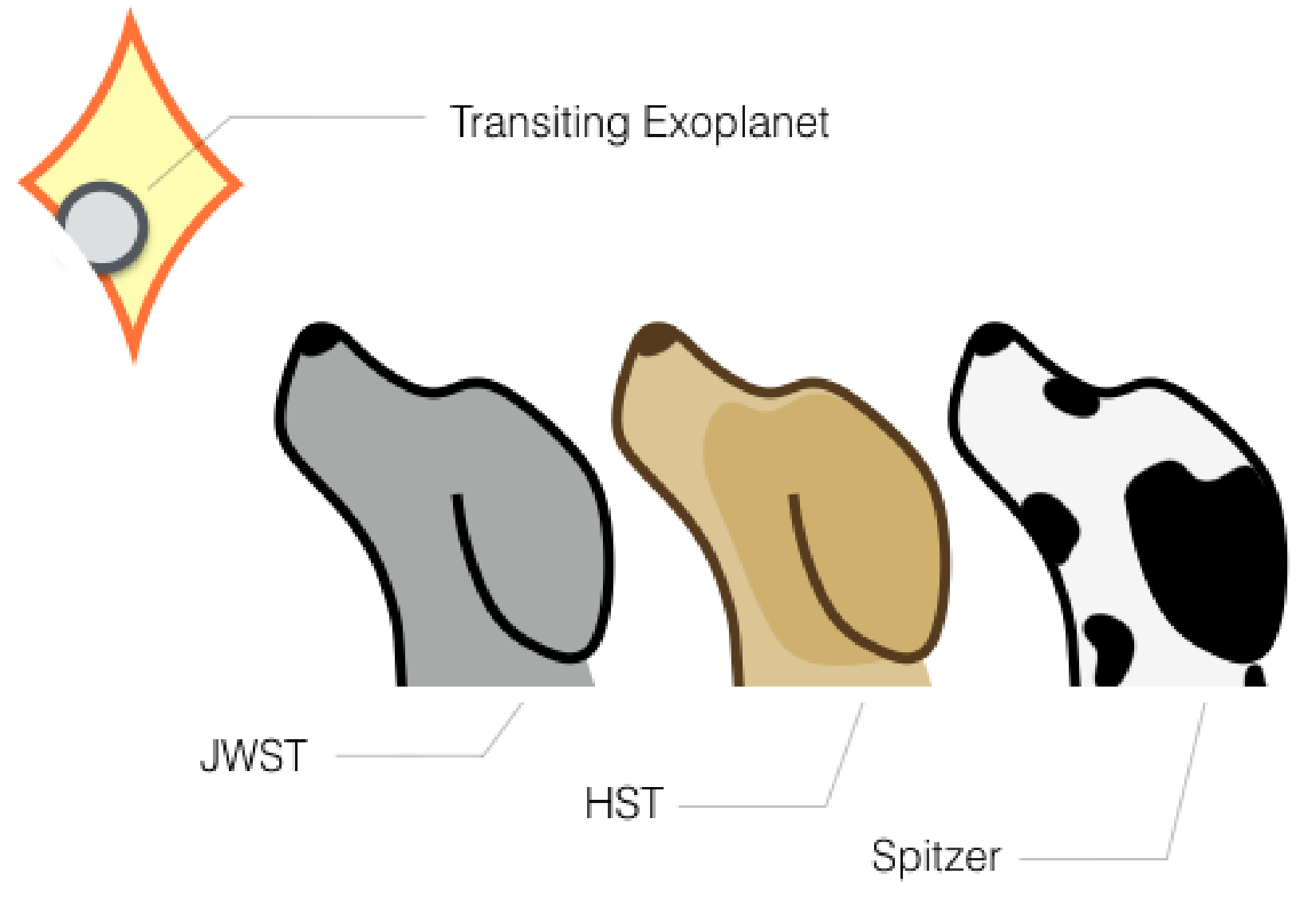}
\caption{
The {\puppies} package is a general TSO analysis tool intended to
offer stress relief as well as a modular capability to process a
multitude of instruments ({\JWST}, {\HST}, {\Spitzer}, etc.).
{\puppies} implements some of the most powerful Machine Learning
algorithms known in the field of Artificial Intelligence, such as
Markov-chain Monte Carlo techniques.
}
\label{fig:pup_diagram}
\end{figure}

Cases like this resonate in individuals pursuing careers in
competitive fields (e.g., the arts or sciences) due to the high
pressure to perform and the power imbalance between members of
such communities.  The philosophical and psychological struggle
to come to terms with academic life can culminate in a
overpowering feeling of the nausea \citep{Sartre1938bookNausea},
like a moth trapped in a bathtub.  To attempt to mitigate some
of these issues and boost the morale of researchers, here we present
The {\bf Pu}blic {\bf P}hotometry {\bf Pi}peline for {\bf
E}xoplanet{\bf s} (\textsc{puppies}), an open-source package to model
time-series observations (TSO) of transiting exoplanets
(Fig.\ \ref{fig:pup_diagram}).


\section{Hounds of Love}
\label{sec:puppies}

The long-awaited start of operations of the {\Webb} (\JWST),
anticipated to launch for the 2004--2009
period \citep[][]{StockmanMather2001iausNGST}, has highlighted the
need for developing open-source community efforts to analyze
astronomical data.  In this context, the {\puppies} package is an
open-source Python package designed to analyze exoplanet time-series
observations for a wide range of instrumentation
(Figure \ref{fig:pup_diagram}).  At present, {\puppies} enables the
reduction and analysis of {\Spitzer} observations \citep[based
on][]{CampoEtal2011apjWASP12b, NymeyerEtal2011apjWASP18b,
StevensonEtal2012apjSpitzerHD149026b, CubillosEtal2013ApjWASP8b,
CubillosEtal2014apjTrES1, LustEtal2014paspLeastAsymmetry}, and
analysis of {\HST} observations \citep{CubillosEtal2020ajHD209458bNUV,
CubillosEtal2023aaHD189733bSTISnuv}.  Development of {\JWST} analysis
tools is anticipated for the near future (2024--2029 period).

But perhaps the most important feature of {\puppies} is to provide the
tools to incorporate more puppies into the daily life of the common
astronomer (see Section \ref{sec:dreaming}).
The {\puppies} software (Python 3.7+) is available for direct
installation from the Python Package Index (PyPI) as:
\begin{quote}
\texttt{pip install exo-puppies}
\end{quote}

or from a conda environment:
\begin{quote}
\texttt{conda install -c conda-forge exo\_puppies}
\end{quote}

The source code is released under the open-source GNU GPL v2 license
at \href{https://github.com/pcubillos/puppies}
{https://github.com/pcubillos/puppies}.  The documentation can be
found at \href{https://puppies.readthedocs.io}
{https://puppies.readthedocs.io}.

\section{The Dreaming}
\label{sec:dreaming}

\subsection{Dog of the day}
\label{sec:dogday}

\begin{figure}[h]
\centering
\includegraphics[width=\linewidth, clip]{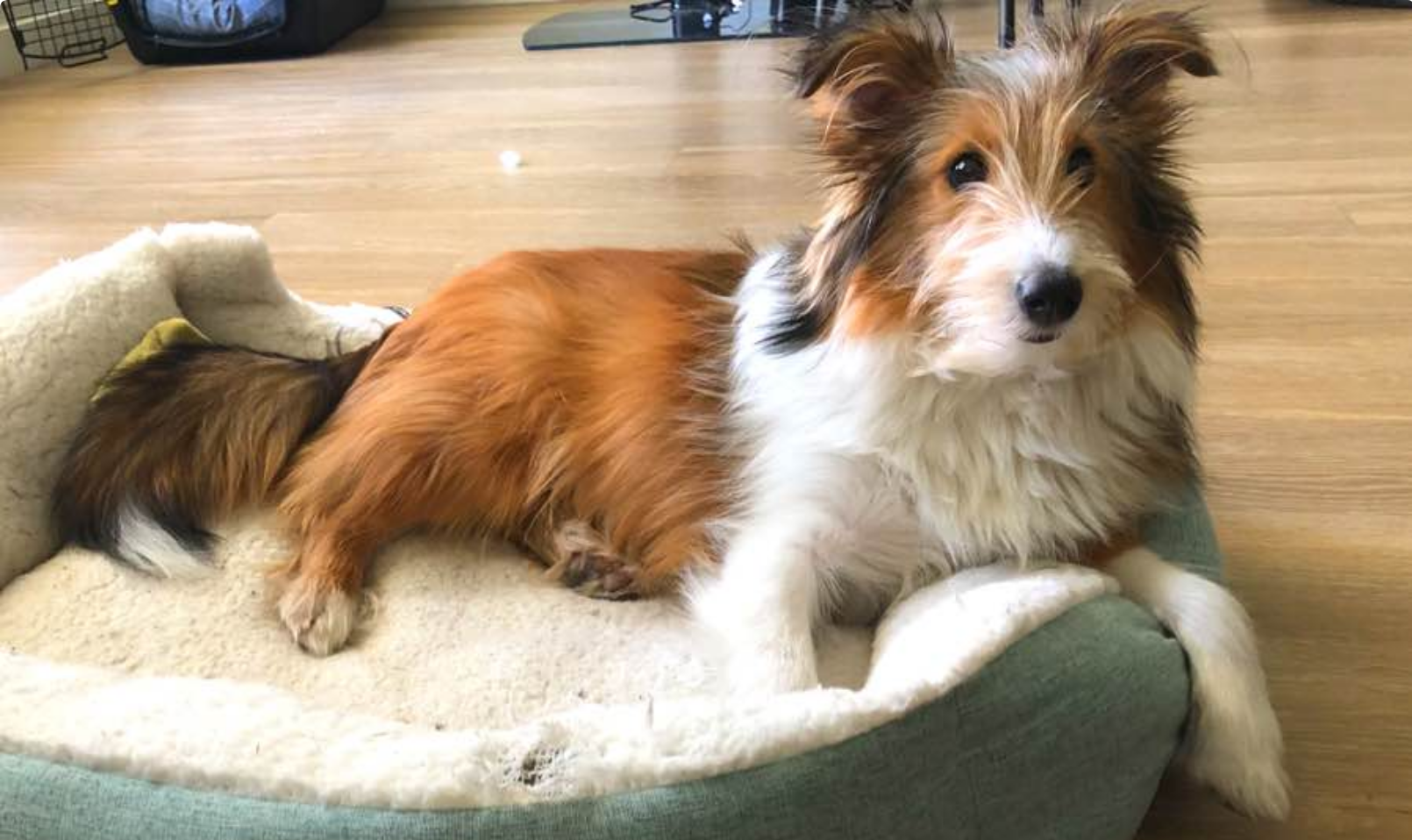}
\caption{
This is Coco (dog of November 28, 2022).  Coco is an eight month old
Sheltie. He is high energy, playful, friendly, and loves to play
fetch. He enjoys relaxing at the feet of his owner.}
\label{fig:dog_day}
\end{figure}

A direct mechanism for stress relief enabled by the {\puppies} package
is the {\it pup of the day}.  Once installed, simply bu executing the
following terminal command, the code will displays on the browser the
dog of the day (for example, see Figure \ref{fig:dog_day}):
\begin{quote}
\texttt{pup {-}{-}day}
\end{quote}


\subsection{Pup matplotlib markers}
\label{sec:matplotlib}

\begin{figure}[tb]
\centering
\includegraphics[width=\linewidth, clip]{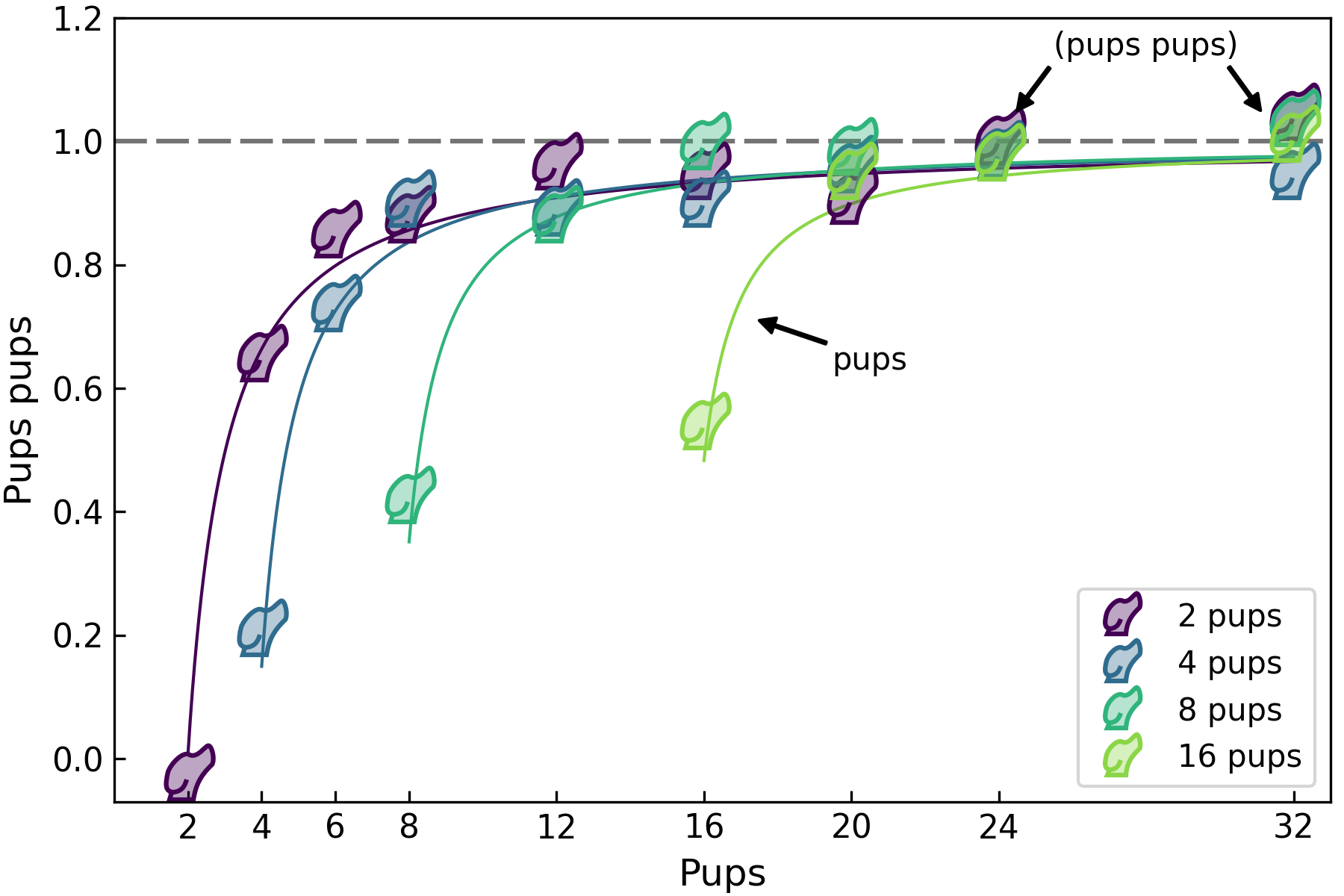}
\caption{Pups pups 2, 4, 8, 16 pups. Pups pups pups pups, pups pups
  (p.p., pups-pups) pups pups \citep[Fig.\ 4 of][]{Zongker2006chicken}.}
\label{fig:chicken}
\end{figure}

Another stress-relief feature provided by the {\puppies} package is
the \texttt{`pup'} marker for matplotlib figures.  The listing
snippet \ref{list:pup} shows how the pup marker can be use in a Python
script.  Figure \ref{fig:chicken} shows a re-edition of Fig.\ 4 of
infamous\footnote{\href{https://isotropic.org/papers/chicken.pdf}{https://isotropic.org/papers/chicken.pdf}}
article by \citet[][]{Zongker2006chicken}.  Who knows, maybe the nicer
marker could be the tipping factor to get your referee to accept your
submission.

\begin{lstlisting}[language=Python, caption=pup marker example., label=list:pup]
import puppies as p
import matplotlib.pyplot as plt

def plotting_with_pups():

    ...

    col = [0, 0, 0, 1.0]
    col_alpha = [0, 0, 0, 0.4]

    fig = plt.figure(0)
    ax = fig.subplot(111)
    ax.plot(
        x, y, marker='pup', ms=8.0, mew=1.5,
        mec=col, mfc=col_alpha,
    )

    ...
\end{lstlisting}

\section{Aerial}
\label{sec:science}

\begin{figure}[tb]
\centering
\includegraphics[width=\linewidth, clip]{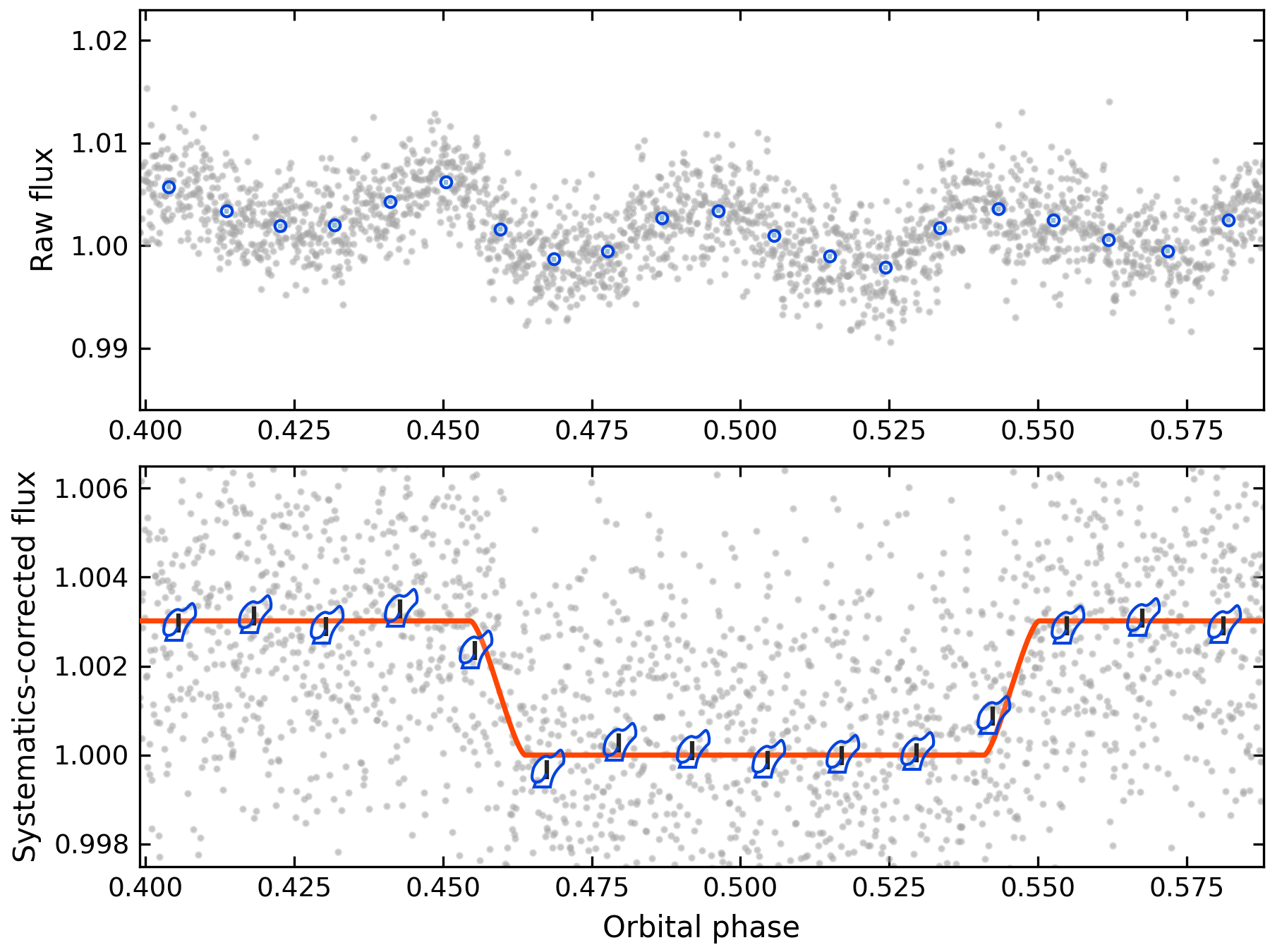}
\caption{WASP-18b secondary eclipse at 3.6 {$\mu$}m observed with {\it Spitzer} (PID 50517). The gray markers represent the individual frames' flux. Blue markers represent the binned flux. The red line shows the eclipse modulation.}
\label{fig:spitzer}
\end{figure}

As an addendum, {\puppies} can also be used to reduce and fit
transiting exoplanet light curves.  Here we show as a sample the
analysis of a {\it Spitzer} secondary eclipse observation of
WASP-18b. For {\it Spitzer}, the {\puppies} pipeline performs the bad
pixel masking, target centering, and aperture photometry to extract
raw light curves.  Then it performs the light-curve modeling by
simultaneously fitting astrophysical and instrumental systematics
parametric models.  The systematics models include both temporal
(ramp) and
pointing \cite[BLISS,][]{StevensonEtal2012apjSpitzerHD149026b} models.
Multiple centering, photometry, and light-curve models settings can be
run and compared \citep[via Bayesian Information Criterion,
BIC,][]{Schwarz1978anstaBIC}.  Figure \ref{fig:spitzer} shows the
{\puppies} reduction of the WASP-18b observation at 3.6 {$\mu$}m.  A
step-by-step script to replicate this analysis can be found in the
{\puppies} documentation\footnote{\href{https://puppies.readthedocs.io/en/latest/WASP18b_eclipse.html}{https://puppies.readthedocs.io/en/latest/WASP18b\_eclipse.html}}.

\section{Never for Ever}
\label{sec:conclusions}

We have presented the {\puppies} package, a tool designed to improve
the quality of research-life for astronomers.  To do so, {\puppies}
brings more puppies into the regular day at the office, either as
pictures of pups or pup markers.  It might also even be useful as an
analysis tool for exoplanet time-series analyses.

\paragraph{Disclaimer}
\label{sec:disclaimer}

This is a work of satire that touches on delicate topics like mental
health, academic stress, or abuse of power; but let's be absolutely
clear that these issues are no joke.  As it was clearly illustrated by
the 2021 EAS Special Session on Welfare and Mental Health in Astronomy
Research\footnote{\href{https://eas21welfaresession.github.io}{https://eas21welfaresession.github.io}},
societal problems like harassment, abuse, and exploitation are not
absent from academic researchers in astronomy.

In the USA, the rate of academic attrition (premature end of their
degree) by doctoral students is between 40\% and
70\% \citep{Gardner2007heTheGrapeVine}.  The relationship between
students and supervisors is the primary reason for student
attrition \citep{DevosEtal2017ejpePhDattrition,
RiglerEtal2017DoctoralAttrition}.  The survey
by \citet{JacobEtal2018AVETHsurvey} found that 24\% of doctoral
students respondents experienced abuse of power from their PhD
supervisors.
Evidence suggests that the rates of
bullying \citep{Einarsen1999ijpWorkplaceBulling} are higher in
academic settings than in other workplace
environments \citep{Moss2018natBullyAcademia}.
\citet{Lovitts2001LeavingAcademie} highlights that
``The most important reason to be concerned about graduate student
attrition is that it can ruin individuals' lives''.

A significant power imbalance exists between PhD students and their
supervisors.  Graduate students are knowledgeable, bright, and
inexpensive labor, which makes them ideal targets of potential
exploitation by their PhD
supervisors \citep{Mendoza2007jheAcademicCapitalism,
MartinEtal2014GraduateMobbing}.  Although many PhD students have
supportive supervisors, but when the relationship with their supervisor
sours, students may find themselves with little support, often feel
powerless \citep{Devlin2018, GoldeDore2001DoctoralExperiences}.
There is limited research on the role that PhD supervisors should play
in doctoral
training \citep{BeginGearard2013pfeSupervisorsRole}. \citet{CohenBaruch2021jbeAbuseExploitation}
developed a model to explain the determinants and outcome of abuse and
exploitation of PhD students, which we summarize up next.

Firstly, doctoral students' performance is harmed by supervisors'
toxic behavior, which may be rooted in the supervisors' struggle to
achieve their goals \citep{Kiley2019aurDoctoralSupervisoryQuality}.
Goals are critical for individuals' self-image and success, so when
they are blocked, PhD supervisors may be motivated to abuse their PhD
students \citep[e.g., by not giving credit for publications or by
requesting research work beyond the PhD
thesis][]{YamadaEtal2014teppWorkplaceBullying}.  Of course, not all
PhD supervisors abuse students, supervisors who abuse or exploit
students {\it choose} to do so by their own
volition \citep{KrasikovaEtal2013jmDestructiveManagement}.

Secondly, the propensity of supervisors to engage in toxic behaviors
can be influenced by their perceptions of their university's ethical
culture \citep{BaskinEtal2015jbamEthicalDisengagement}.  Organizations
are considered ethical when their members show greater concern toward
the interests of others than toward their own interest.  In the
absence of supportive ethical cultures, PhD supervisors will believe
that norms of fairness are not enforced; encouraging them to exploit
their students \citep{HsiehWang2016EthicalClimateDeviance}.  Moreover,
in an abusive culture the abuse might actually be perceived to be
normal, or even positive.

Lastly, PhD supervisors' with tendencies to emphasize self-interest
over the interests of others, they will be more inclined to engage in
abuse and exploitation of their doctoral
students \citep{KrasikovaEtal2013jmDestructiveManagement}.
Individuals characterized by Dark Triad personality (DTP) traits display the most
prominent disposition toward
self-interest \citep{KrasikovaEtal2013jmDestructiveManagement}.  The
Dark Triad consists of three anti-social personality traits:
manipulation of others, narcissism (feelings of entitlement and
superiority), and psychopathy accompanied by low empathy and
anxiety \citep{OBoyleEtal2011gomBadApplesBadBarrels}.

Institutions play a significant role in shaping in how members of
academia behave.  When policies and sanctions exist and are enforced,
less abuse and exploitation of students by DTPs is expected.
Supervisors will hesitate before performing abusive and exploitative
behavior when they know their students are fully aware of their
rights.
Overcoming these challenges require the active initiative of all
members of the community, including those in a position of privilege;
silence may indirectly encourage abuse and
exploitation \citep{CyranoskiEtal2011natPhDfactory}.  Simply taking a
moment to reflect can improve ones perspective; as put in
retrospective by J. Timberlake regarding B. Spears: ``I do not want to
ever benefit from others being pulled
down\footnote{\href{https://www.bbc.com/news/entertainment-arts-56047830}{https://www.bbc.com/news/entertainment-arts-56047830}}.''

\acknowledgments

We thank contributors to the Python Programming Language and the free
and open-source community (see Software Section below).  We drafted
this article using the AASTeX6.2 latex
template \citep{AASteamHendrickson2018aastex62}, with further style
modifications that are available
at \href{https://github.com/pcubillos/ApJtemplate}
{https://github.com/pcubillos/ApJtemplate}.  This work is based in
part on observations made with the {\SST}, which is operated by the
Jet Propulsion Laboratory, California Institute of Technology under a
contract with NASA.  This research has made use of NASA's Astrophysics
Data System Bibliographic Services.

{\color{white} Hello, this is dog!}

\software{\\
{\puppies} \citep{Cubillos2024aprilPuppies},
\textsc{mc3} \citep{CubillosEtal2017apjRednoise},
\textsc{Numpy} \citep{HarrisEtal2020natNumpy},
\textsc{SciPy} \citep{VirtanenEtal2020natmeScipy},
\textsc{Astropy} \citep{AstropyCollaboration2013aaAstropy,
 AstropyCollaboration2018ajAstropy},
\textsc{Matplotlib} \citep{Hunter2007ieeeMatplotlib},
\textsc{IPython} \citep{PerezGranger2007cseIPython},
AASTeX6.2 \citep{AASteamHendrickson2018aastex62},
and
\textsc{bibmanager}\footnote{
\href{http://pcubillos.github.io/bibmanager}
     {http://pcubillos.github.io/bibmanager}}
\citep{Cubillos2019zndoBibmanager}.
}

\bibliography{puppies_aprilis}

\appendix

\section{Receipt}
\label{sec:resort}

\begin{figure}[h]
\centering
\includegraphics[width=\linewidth, clip, trim=0 500 540 0]{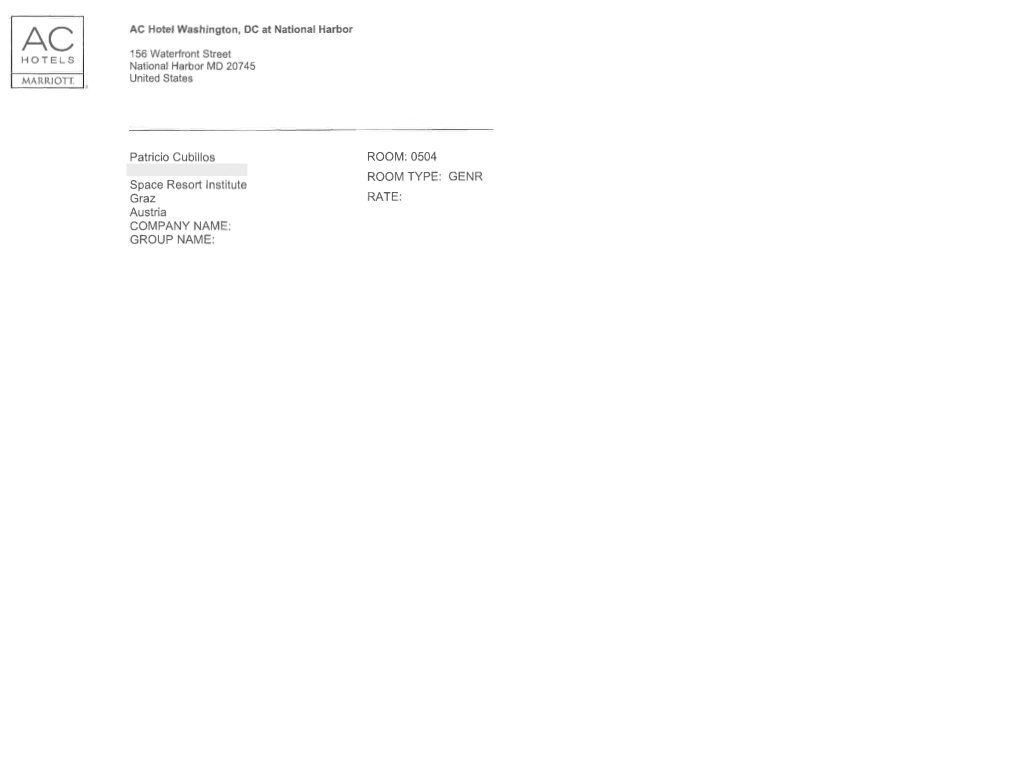}
\caption{Marriot hotel receipt.}
\label{fig:marriot}
\end{figure}

Figure \ref{fig:marriot} shows the actual receipt from that one
receptionist at Marriot Hotel.

{\color{ghost}

\section{Pop-quiz time}
\label{sec:quiz}

The first 15 people who can tell me, without googling, what is the
link between the section names will get a free {\puppies} sticker.

}

\end{document}